\documentclass[pre,preprint,showpacs,preprintnumbers,amsmath,amssymb]{revtex4}

\usepackage{ulem}
\newcommand{\asja}[1]{{#1}}
\begin{document}

\title{Dissipative electromagnetism from a nonequilibrium thermodynamics perspective}
\author{Asja Jeli\'c}
\author{Markus H\" utter}

%\email[Corresponding Author: Address:\\
%ETH Zurich\\
%Department of Materials, Institute of Polymers\\
%HCI H 529\\
%Wolfgang-Pauli-Str. 10 \\
%8093 Zurich\\
%SWITZERLAND \\
%Phone: +41 44 632 45 44\\
%Fax:   +41 44 632 10 76\\
%Email: markus.huetter@mat.ethz.ch]{}

\author{Hans Christian \"Ottinger}
\affiliation{Department of Materials, Institute of Polymers, ETH
Zurich, 8093 Zurich, Switzerland.}

\date{\today}

\begin{abstract}
Dissipative effects in electromagnetism on macroscopic scales are
examined by coarse graining the microscopic Maxwell equations with
respect to time. We illustrate a procedure to derive the
dissipative effects on the macroscopic scale by using
a Green-Kubo type expression in terms of the microscopic
fluctuations and the correlations between them.
The resulting
macroscopic Maxwell equations are formulated within the General
Equation for the Non-Equilibrium Reversible-Irreversible Coupling
(GENERIC) framework, accounting also for inhomogeneous
temperature.
\end{abstract}


\maketitle

\newpage

\section{Introduction} \label{sec1}

\hspace*{\parindent}

\asja{The dynamics of matter can be described on different levels
of detail. For example, microscopic descriptions resolve short
length scales and fast processes. In principle, one can follow the
microscopic dynamics over very long times to arrive at the
behavior on the macroscopically relevant time scales. However,
doing so is often not desirable, and one can learn significantly
more by using a coarse-grained description. For example, the flow
of water in a complex geometry is usually studied not in terms of
the dynamics of water molecules, but rather in terms of the
continuum equations of hydrodynamics. This is so because key
features of interest emerge only on the length and time scales
much larger than the molecular ones. The intrinsic time scale for
the relaxation of inhomogeneities in the macroscopic velocity
field is closely related to the shear viscosity $\eta$. This
material specific property can either be determined experimentally
on purely macroscopic grounds. Another, conceptually very
interesting, route goes via the so-called fluctuation-dissipation
theorem }\cite{Kubo,Groot,Evans,hco}\asja{, which establishes a
relation between micro- and macroscopic properties. In particular,
the Green-Kubo formula relates the shear viscosity to the
correlations between fluctuations of the microscopic shear stress,
$\sigma_{x y}$, by way of}
\begin{eqnarray}\label{viscosity}
\eta &=& \frac{V}{k_{\rm B}T}\int\limits_0^{\infty} \langle
\sigma_{x y}(t)\sigma_{x y}(0) \rangle_{\mathrm{eq}} \, dt,
\end{eqnarray}
\asja{with $V$ the sample volume, $T$ the absolute temperature,
and $\langle \ldots \rangle_{\mathrm{eq}}$ the equilibrium
average. The integration range $[0, \infty]$ stands symbolically
for an integration range much longer than any microscopic
relaxation time. However, it is implicitly assumed that this range
is shorter than macroscopic time scales, which implies that micro-
and macroscopic timescales are clearly separated. Relations of
this type are frequently used in molecular dynamics simulations to
determine macroscopic transport properties \cite{Evans}, where
transport coefficients are expressed in terms of two-time
correlations of current densities.}

\asja{The motivation for the work presented here related to the
  Maxwell equations rests on the
following idea. The Green-Kubo relation \eqref{viscosity}, and
generalizations thereof, not only tells us {\it how} to relate
microscopic fluctuations to macroscopic transport coefficients.
More fundamentally, it indicates {\it that} microscopic
fluctuations give rise to dissipative processes in more
coarse-grained descriptions. In view of electromagnetism, this
serves as a strong motivation to believe that, e.g., fluctuations
in the polarization and magnetization by way of particle
vibrations and fast spin dynamics must be observable on
macroscopic scales as dissipative processes. Although the Maxwell
equations belong to the most studied differential equations in
physics, they do not account for the dissipative phenomena beyond
Ohmic resistance and the frequency dependent imaginary parts of
the permittivity and permeability in the linear response regime.
In view of the above discussion, we anticipate that the electric
(Ohmic) resistance is related to the fluctuations of the electric
current of the unbound charges. However, in addition also the
bound particles fluctuate, giving rise to fluctuations in the
polarization and the magnetization. How those fluctuations give
rise to dissipative effects on the
  macroscopic scale is
precisely what we want to elaborate on in this
manuscript.}

\asja{Thermodynamics comes into play in the Green-Kubo relation
(\ref{viscosity}) by way of the absolute temperature $T$. In
addition, the occurrence of a ``thermal'' variable is also of
fundamental importance in a purely macroscopic model with
dissipative effects, since the latter lead to an entropy increase,
and in turn to a change in the thermal state. In contrast, the
Maxwell equations are usually taken as a set of isolated
equations, decoupled from the other macroscopic variables, e.g.,
temperature and density. This means that the evolution of the
electromagnetic field is generally obtained in the approximation
that these macroscopic variables have set values or, at best, are
given as functions of time. In order to obtain a rigorous and
reliable theory, one needs to include these macroscopic quantities
in the list of variables and to consider their dynamics in
conjunction with the time evolution of the electromagnetic field.
Only then one can capture the interplay between the thermodynamic
behavior and the electromagnetic field, and dissipative effects in
electromagnetism can be addressed in a consistent manner.}

To perform a thorough theoretical analysis of the interplay
between the thermodynamic properties and the electromagnetic field
is a demanding task, especially if one wants to account for
dissipation effects beyond a linear response theory. Liu and
coworkers have incorporated dissipative effects into the Maxwell
equations \cite{Liu93,Liu94,Liu95,Liu98,Liu99,Liu02,Jiang96}. The
resulting theory describes the dynamics of macroscopic systems
that are exposed to electromagnetic fields and contain electric
charges and currents. \asja{In particular, the driving forces
for the dissipative effects on the macroscopic scale have been
established. In the dynamic equations, the new dissipative
terms occur as an addition to the commonly used macroscopic
Maxwell equations} when the system is out of equilibrium. The
validity of this theory was at first confined to the low-frequency
regime, but was later generalized to higher frequencies
\cite{Jiang96}.

In this paper the goal is to use the reversible Maxwell equations
to derive macroscopic Maxwell equations for dissipative
electromagnetism in a medium, valid also outside of the linear
response regime. As the guideline to complete this task, the
General Equation for the Non-Equilibrium Reversible-Irreversible
Coupling (GENERIC) formalism of non-equilibrium thermodynamics
\cite{hco,hco1,hco2} is used \asja{for two reasons. First,
thermodynamic consistency of the macroscopic model is ensured. And
second, the formalism is equipped with a scheme for temporal
coarse-graining, the most important part of which is concerned
with the emergence of dissipative effects upon coarse-graining,
i.e., a generalization of the Green-Kubo relation
\eqref{viscosity}. This scheme will be illustrated below
specifically on the Maxwell equations} in order to obtain the
macroscopic equations that possess the GENERIC structure. These
equations are then compared to the dissipative Maxwell equations
suggested by Liu \cite{Liu93}.

The paper is organized as follows. In Sec.~\ref{sec2}, the
microscopic and common macroscopic Maxwell equations are briefly
presented. The usual coarse-graining procedure and its
shortcomings are recapitulated. In Sec.~\ref{sec3}, a GENERIC
formulation of the Maxwell equations together with the
equation for \asja{the energy density} is presented. We
illustrate the procedure to obtain, by temporal coarse-graining
\asja{using Green-Kubo-type expressions, the dissipative effects
in} the time evolution equations for the macroscopic variables.

\section{Micro- and macroscopic Maxwell equations} \label{sec2}

The classical derivation of the macroscopic Maxwell equations for
the electromagnetic fields in a medium from the microscopic ones,
\begin{subequations} \label{micro}
\begin{eqnarray}
\dot {\mathbf b}&=& -\nabla \times {\mathbf e} \ , \qquad \qquad
\, \qquad \qquad \nabla \cdot {\mathbf b}= 0 \ , \label{micro_b}
\\ \dot {\mathbf e}&=&\frac{1}{\epsilon_0 \mu_0}\nabla \times
{\mathbf b} -\frac{1}{\epsilon_0} {\mathbf j} \ , \qquad \qquad
\nabla \cdot {\mathbf e}= \frac{1}{\epsilon_0} \rho \
\label{micro_e},
\end{eqnarray}
\end{subequations}
is often performed through spatial averaging \cite{Jackson}.
Throughout the manuscript we denote the partial time derivative by
$\dot {\mathbf A} \equiv \frac{\partial}{\partial t} {\mathbf A}$. The
quantities $\epsilon_0$ and $\mu_0$ stand for the permittivity and
the permeability, respectively, in vacuum. Starting from the
microscopic Maxwell equations containing only fields $\mathbf e$
and $\mathbf b$, one separates the free from the bound parts of
the charge $\rho$ and of the current density $\mathbf j$. Upon
spatial averaging, the expressions for the bound parts of $\rho$
and $\mathbf j$ are simplified by introducing the polarization
${\mathbf P}$ and the magnetization ${\mathbf M}$. With
$\mathbf{E}$  and $\mathbf{B}$ the spatially averaged electric
field and magnetic induction, respectively, the macroscopic
Maxwell equations can then be formulated conveniently as
\begin{subequations} \label{macro}
\begin{eqnarray}
\dot {\mathbf B}&=& -\nabla \times {\mathbf E} \ , \qquad \qquad
\qquad \nabla \cdot {\mathbf B}=0 \ , \label{Maxwell_B}\\ \dot
{\mathbf D}&=&\nabla \times {\mathbf H}-{\mathbf J}_e \ , \qquad
\qquad \ \nabla \cdot {\mathbf D}=\rho_e \ , \label{Maxwell_D}
\end{eqnarray}
\end{subequations}
where one has introduced the electric displacement ${\mathbf
D}=\epsilon_0 {\mathbf E}+{\mathbf P}$, assuming that quadrupol
moments are negligible, and the magnetic field ${\mathbf
H}={\mathbf B}/\mu_0-{\mathbf M}$. Doing so, the source terms of
the macroscopic Maxwell equations are given by the averages of the
free charge density and free current density, $\rho_e$ and ${\mathbf J}_e$,
respectively. In addition, the evolution equation for the energy density
of the electromagnetic field
can be written in the form
\begin{eqnarray}\label{energy_equation}
\dot{\varepsilon}_{\rm em}&=&
- \nabla \cdot\left({\mathbf E}\times {\mathbf
H}\right)-{\mathbf J}_e\cdot {\mathbf E},
\end{eqnarray}
where ${\mathbf E}\times {\mathbf H}$ is the Poynting vector.

The structure of the equations \eqref{macro} imposes the
interpretation that the field variables are $\mathbf D$ and
$\mathbf B$, the temporal equations are their time evolution
equations, while the other two represent the constraints that must
be satisfied at all times. The fields ${\mathbf E}$ and ${\mathbf
H}$ then need to be expressed as some functions of $\mathbf D$ and
$\mathbf B$ in order to render equations \eqref{macro} closed.
Such closing relations are known as constitutive equations,
which contain the information about the medium, and are usually
considered as given in macroscopic electrodynamics. The usual
choice is to use linear relations between $(\mathbf E,\mathbf B)$
and $(\mathbf D,\mathbf H)$, valid if the electromagnetic field is
sufficiently weak, in order to obtain the properties of the
permittivity $\mbox{ \boldmath $ \epsilon$}$ and permeability
$\mbox{ \boldmath $\mu$}$ that describe the response of a medium
on the electromagnetic field. Then the real parts of the functions
$\mbox{\boldmath $\epsilon $}(\omega)$ and $\mbox{\boldmath$
\mu$}(\omega)$ express the oscillatory motion of the bound charges
and the imaginary parts describe dissipation. However, in many
cases, like the passage of the laser beam through a substance, or
for a medium in a strong magnetic field, the linear response
theory is not valid anymore and one needs to work in the regime of
nonlinear electrodynamics, with the consequence of loosing all the
simple relations.

\asja{A common constitutive relation concerns the current density.
%In view of studying the Maxwell equations from a
%nonequilibrium thermodynamics perspective, we draw the following
%analogy.
It is known that Ohmic conductors satisfy the
condition ${\mathbf E}^{\rm eq}={\mathbf 0}$ at equilibrium, and that
in non-equilibrium situations the force to relax
${\mathbf D}$ to equilibrium is proportional to
${\mathbf E}$. Such behavior is often expressed by
${\mathbf J}_e = \sigma {\mathbf E}$ with conductivity $\sigma$.
This interrelation suggests to draw the following analogy.
In stationary, equilibrium situations,
with vanishing electric current ${\mathbf J}_e$, one obtains from
\eqref{macro} the conditions}
\begin{eqnarray}\label{thd_forces}
\nabla \times {\mathbf H}^{\rm eq}={\mathbf 0} \ ,  \ \ \nabla
\times {\mathbf E}^{\rm eq}={\mathbf 0} \ .
\end{eqnarray}
\asja{We therefore assume
that in non-equilibrium situations on the macroscopic scale
there are forces
to drive $\mathbf B$ and $\mathbf D$ towards equilibrium with those
forces being
closely related to $\nabla \times {\mathbf H}$ and $\nabla
\times {\mathbf E}$ with associated transport coefficients.
Conditions \eqref{thd_forces} have been
obtained also by Liu by minimizing the total energy of the system
under the constraints of the non-temporal Maxwell equations
$\nabla \cdot {\mathbf D}=\rho_e$ and $\nabla \cdot {\mathbf B}=0$
\cite{Liu93}.}

The commonly used macroscopic Maxwell equations \eqref{macro} can
be obtained not only by spatial averaging (see e.g.\
\cite{Jackson}), which is also called ``truncation'' in
\cite{Rob}, but also by ensemble averaging \cite{Rob, GrootSutt}.
In both cases, only single time properties are considered, while
temporal correlations and two-time ensemble averages are
neglected. In the remainder of this manuscript, we will thus refer
to these two techniques as ``spatial'' averaging, in contrast to
temporal coarse-graining which will be discussed in detail below.
Neglecting coarse-graining with respect to time to arrive at the
spatially averaged Maxwell equations  \eqref{macro} may be a
valuable ansatz under many circumstances \cite{Jackson}. However,
it is also known that coarse-graining with respect to space {\it
and} time can lead to a better understanding, and some aspects of
a model emerge clearly only on longer time scales \cite{hco,oett}.
\asja{Such an example is hydrodynamics, where the shear viscosity
given by \eqref{viscosity} and other transport coefficients in the
Navier-Stokes equations arise at the continuum level of
description, but are absent in a molecular model.} Therefore, we
will examine specifically the effect of coarse-graining in time
\asja{by using Green-Kubo-type expressions} to learn more about
dissipative effects in electromagnetic systems \asja{with
nonlinear constitutive relations. In order to obtain a closed set
of the macroscopic time evolution equations for electromagnetism
in a medium, a scheme for temporal coarse-graining of the
``spatially averaged'' Maxwell equations is presented following
the GENERIC formalism} \cite{hco,hco1,hco2}. The full
coarse-graining procedure together with the expressions obtained
for \asja{the additional dissipative terms} is presented in the
following sections.

\section{GENERIC formulation} \label{sec3}

\subsection{Formalism}

\asja{The main points of the GENERIC framework of nonequilibrium
thermodynamics} \cite{hco,hco1,hco2} \asja{can be summarized
briefly in the following way. In analogy to equilibrium
thermodynamics, major importance comes to, first, choosing a
complete set of variables $\mathbf{x}$, which describes the
situation of interest to the desired detail. The reversible
contributions to the time evolution for this set of variables,
$\dot {\mathbf x}|_{\rm rev}$, is formulated in close reference to
classical Hamiltonian mechanics. In particular, it is related to
the energy gradient by way of a Poisson operator ${\mathbf L}$,
i.e., $\dot {\mathbf x}|_{\rm rev} = {\mathbf L} \cdot (\delta
E/\delta {\mathbf x})$. The Poisson bracket associated to
${\mathbf L}$, given by $\{A,B\} = ( \delta A/\delta {\mathbf x},
{\mathbf L} \cdot \delta B/\delta{\mathbf x})$ with appropriate
scalar product $( \ldots )$, must be antisymmetric and satisfy the
Jacobi identity, which are both abstract features that capture the
nature of reversibility. In order to formulate the irreversible
part of the dynamics one is motivated by the reversible part. For
the reversible dynamics, the energy plays a distinct role. First,
it is a conserved quantity for closed systems and, second, it
drives the reversible dynamics. In parallel, for irreversible
dynamics entropy is a fundamentally important quantity, which must
not decrease for closed systems. In analogy to the reversible
dynamics, it is assumed in the GENERIC framework that the
irreversible contributions to the time evolution of $\mathbf{x}$,
$\dot {\mathbf x}|_{\rm irrev}$, is driven by the entropy
gradient, i.e., that it is of the form $\dot {\mathbf x}|_{\rm
irrev} = {\mathbf M} \cdot (\delta S/\delta {\mathbf x})$, with
${\mathbf M}$ a generalized friction matrix. The latter is
required to be (Onsager-Casimir) symmetric, and contains transport
coefficients and relaxation times associated to the corresponding
dissipative effects. The condition that the friction matrix is
positive semi-definite ensures that $\dot S \ge 0$ is fulfilled.}

\asja{In summary, the time evolution
of $\mathbf{x}$ can be expressed in terms of four building
blocks $E$, $S$, $\mathbf L$ and $\mathbf M$ as}
\begin{equation}\label{generic}
%\frac{{\text d}\mathbf{x}}{{\text d}t}
\dot {\mathbf x}
={\mathbf
L}(\mathbf{x})\cdot \frac{\delta E(\mathbf{x})} {\delta
\mathbf{x}}+{\mathbf M}(\mathbf{x})\cdot \frac{\delta
S(\mathbf{x})}{\delta \mathbf{x}} \; .
\end{equation}
\asja{The two different contributions to the time
evolution, reversible and irreversible, are not independent. Rather,
they are interrelated by the two degeneracy requirements}
\begin{equation}\label{degeneracy}
{\mathbf L}(\mathbf{x})\cdot \frac{\delta S(\mathbf{x})}{\delta
\mathbf{x}}=\mathbf{0} \; , \; \; \; {\mathbf M}(\mathbf{x})\cdot
\frac{\delta E(\mathbf{x})}{\delta \mathbf{x}}=\mathbf{0} \; .
\end{equation}
\asja{The first condition expresses the reversible nature of the
$\mathbf L$ contribution to the dynamics, demonstrating the fact
that the reversible dynamics captured in $\mathbf L$ does not
affect the entropy functional. The second one expresses the
conservation of the total energy of an isolated system by the
irreversible contribution to the system dynamics captured in
$\mathbf M$.}

\asja{A particular feature of the GENERIC is that it is applicable
on different levels of description, including reversible
Hamiltonian mechanics and dissipative macroscopic field theories.
The corresponding four building blocks obviously differ between
the different levels. However, there are abstract procedures to
relate them. Most importantly, the friction matrix on a
coarse-grained level can be expressed in terms of the more
microscopic dynamics \cite{hco}. This procedure helps to
understand how dissipative effects can arise in coarse-grained
descriptions. It is particularly this issue that we want to
address in relation to the Maxwell equations of electrodynamics.}

\subsection{Choice of variables, energy and entropy functionals}

First, we identify a natural set of variables for a dynamic
description of dissipative electromagnetism. In order to describe
a body in an electromagnetic field, we make the local equilibrium
assumption, so that temperature, internal energy and entropy
densities are defined at each point of the nonequilibrium system.
\asja{Although one is free to choose any of these three
variables for describing the ``thermal'' state of the body, it
will become clear below that it is particularly useful to choose
the energy density $\varepsilon$, since it is the density of a variable
which is conserved for a closed system. The energy density
$\varepsilon$ stands for the total energy density inside the
system volume, consisting of both the internal energy of matter and the
energy of the electromagnetic field.}
As far as the variables describing the electromagnetic
state are concerned, we keep in mind that our primary interest is
in the effect of coarse-graining the Maxwell equations with
respect to time. Hence, we use the \asja{already} spatially
averaged fields $\mathbf D$ and $\mathbf B$ as used in the
temporal Maxwell equations in \eqref{macro}.
\asja{Since our focus here is on the temporal coarse-graining
and on constructing the dissipative effects in a macroscopic
formulation of electromagnetism, we consider only bodies at rest
to avoid additional complexity, i.e., we do not include the
velocity field in the set of variables. This restriction, in turn,
also means that the mass density is a constant, as one can easily
infer from the mass balance equation. In other words, we restrict
our attention} to materials for which the volume changes are
negligible compared to the other effects of interest, namely
materials with vanishingly small isothermal compressibility and
thermal expansion coefficient. \asja{The full set of variables is
thus given by ${\mathbf x}=\left( \varepsilon,{\mathbf D},{\mathbf
B} \right) $. }The total energy and entropy of the system,
expressed in terms of these variables, are given by the
functionals
\begin{subequations}
\begin{eqnarray}
E &=& \int \varepsilon  \ d^3r  \, ,\\
S &=& \int s\left( \varepsilon,{\mathbf D},{\mathbf B} \right) \
d^3r \ .
\end{eqnarray}
\end{subequations}
\asja{The functional derivatives of $E$ and $S$ will be needed
  further below for the calculation of time evolution of
variables $\mathbf x$ by using the general evolution equation
\eqref{generic}.}
\asja{In order to
connect the expressions for these functional derivatives} to known
quantities, one refers to the Gibbs thermodynamic relation. The
latter gives the change in the energy density $\varepsilon$ of a
thermally isolated medium at rest in the form of the total differential
\begin{equation}\label{Gibbs}
d\varepsilon=Tds +{\mathbf E}\cdot d{\mathbf D}+{\mathbf H}\cdot
d{\mathbf B}\, ,
\end{equation}
under the assumption of local equilibrium,\asja{ where  the fields
${\mathbf E}$ and ${\mathbf H}$, as well as the temperature $T$,
are given as the partial derivatives of the energy density
$\varepsilon$ with respect to the appropriate variables. The first
term on the right side in \eqref{Gibbs}} is identical to the
ordinary thermodynamic relation for the internal energy density in
the absence of the electromagnetic field and for a constant mass
density. The last two terms represent the change of the energy
density due to changes in the electromagnetic fields. The form of
these two terms is well known from the standard textbooks on
electrodynamics \cite{Jackson,Landau}, where the total
electromagnetic energy increment is given as a volume integral
over the increments $\delta {\mathbf D}$ and
$\delta {\mathbf B}$ multiplied by ${\mathbf E}$ and
${\mathbf H}$, respectively. Relation \eqref{Gibbs} is also in
agreement with the total energy density used in \cite{Feld99} for
a system at rest. We point out that
the precise form of \asja{the energy
density $\varepsilon$ of the system} is not required below for the
derivation of the macroscopic Maxwell equations in the GENERIC
form.

\asja{The functional derivatives $\delta E/\delta{\mathbf x}$
and $\delta S/\delta{\mathbf x}$ are now obtained by reordering
Eq.\eqref{Gibbs} in terms of the entropy total differential, and
calculating the appropriate partial derivatives therefrom. One
finds the following expressions needed for the GENERIC time
evolution equation \eqref{generic}:}
\begin{subequations} \label{funcder}
\begin{eqnarray}
\frac{\delta E}{\delta {\mathbf x}} &=& \left(
\begin{array}{c}
1 \\ {\mathbf 0} \\ {\mathbf 0}
\end{array}
\right) , \label{energy_derivative} \\ \frac{\delta S}{\delta
{\mathbf x}} &=& \frac{1}{T} \left(
\begin{array}{c}
1 \\
-{\mathbf E}
\\ -{\mathbf H}
\end{array}
\right) . \label{entropy_derivative}
\end{eqnarray}
\end{subequations}

\subsection{\asja{Reversible dynamics}}

\asja{The reversible contribution to the time evolution of the
variables
${\mathbf x}=\left( \varepsilon,{\mathbf D},{\mathbf B} \right)$ is
related to the energy gradient $\delta E/\delta {\mathbf x}$ by way of
the Poisson operator ${\mathbf L}$.
At this stage, we point out that the effects arising through
coarse-graining in time by using the Green-Kubo formula are
completely comprised in the friction matrix $\mathbf M$, i.e., in
the irreversible part of the time evolution. Correspondingly, temporal
coarse-graining does not affect
the reversible part of the time evolution of the
variables $\mathbf x$. Therefore, the Poisson operator}
\begin{eqnarray}
{\mathbf L}& =& \left(
\begin{array}{ccc}
L_{\varepsilon \varepsilon} & L_{\varepsilon{\mathbf D}}&
L_{\varepsilon{\mathbf B}}
\\
L_{{\mathbf D}\varepsilon} & L_{{\mathbf D}{\mathbf D}}&
L_{{\mathbf D}{\mathbf B}}
\\
L_{{\mathbf B}\varepsilon}& L_{{\mathbf B}{\mathbf D}} &
L_{{\mathbf B}{\mathbf B}}
\end{array}
\right)
\end{eqnarray}
\asja{can be build up from the already established Maxwell
equations \eqref{macro} and the adequate evolution equation for the
total energy density, given only by the first term in
Eq.\eqref{energy_equation} for a medium at rest.}

\asja{We approach the construction of $\mathbf L$ in the
following manner, having in mind that the reversible contributions
are of the form ${\mathbf L}\cdot \left(\delta E/ \delta
\mathbf{x}\right)$. Since only the first element of the vector
$\delta E/\delta {\mathbf x}$ is non-zero,
Eq.\eqref{energy_derivative}, terms $L_{{\mathbf D}\varepsilon}$
and $L_{{\mathbf B}\varepsilon}$ must reproduce the time evolution
equations \eqref{macro}. The antisymmetry of $\mathbf L$ is then
used to specify the elements $L_{\varepsilon{\mathbf D}}$ and
$L_{\varepsilon{\mathbf B}}$. With this, the
$\varepsilon$-component of the degeneracy requirement ${\mathbf
L}\cdot \left( \delta S/ \delta \mathbf{x}\right)={\mathbf 0}$ can
be satisfied by choosing $L_{\varepsilon \varepsilon}$
accordingly. Notice that this term is then also in consistence
with the time evolution for $\varepsilon$, and it is antisymmetric as
well. The other two components of the degeneracy condition can be
fulfilled by choosing $L_{{\mathbf D}{\mathbf B}}= \nabla \times$,
$L_{{\mathbf B}{\mathbf D}}= -\nabla \times$, and $L_{{\mathbf
D}{\mathbf D}}=L_{{\mathbf B}{\mathbf B}}={\mathbf 0}$. In summary,
one obtains the Poisson operator}
\begin{eqnarray}\label{L-matrix}
{\mathbf L} = \left(
\begin{array}{ccc}
{\mathbf E}\cdot \nabla \times {\mathbf H} -{\mathbf H}\cdot
\nabla \times {\mathbf E} \, \, \, \, \, &-{\mathbf H}\cdot \nabla
\times \, \, \, \, \, & {\mathbf E}\cdot \nabla \times
\\
\nabla \times {\mathbf H} & {\mathbf 0} & \nabla \times
\\
-\nabla \times {\mathbf E} & -\nabla \times & {\mathbf 0}
\end{array}
\right) ,
\end{eqnarray}
\asja{with all derivative operators acting on everything to the
right, i.e., also on functions multiplied to the right side of
the operator $\mathbf L$. At this point we mention that, in general,
the symbol ``$\cdot$'' in \eqref{generic} implies not only summation
over discrete indices. If field variables are involved the
operators
$\mathbf L$ and $\mathbf M$ are written in terms of two space
arguments $\left({\mathbf r},{\mathbf r'}\right)$, and an
integration over ${\mathbf r'}$ must be performed when multiplied with a function of
${\mathbf r}'$ from the right. However,
in the case of the field equations being
local, one can express $\mathbf L$ and $\mathbf M$ in terms of a
single variable
$\mathbf r$ only \cite{hco}, and no integration is implied when these
operators are multiplied from the right. Such single variable notation
is used for the Poisson operator \eqref{L-matrix}. However, in case of
the friction matrix
$\mathbf M$, discussed in the next section, it will be beneficial to
keep the general form in terms of ${\mathbf r}$ and ${\mathbf r'}$.}

\asja{The reversible time evolution of the variables
$\mathbf x$ as obtained from ${\mathbf L} \cdot \delta E/{\mathbf x}$
takes the form}
\begin{subequations} \label{rev_evolution}
\begin{eqnarray}
{\dot \varepsilon}_{\rm rev}&=& -\nabla \cdot \left({\mathbf
E}\times {\mathbf H}\right) ,
\label{eq1} \\
\dot{\mathbf D}_{\rm rev} &=& \nabla \times {\mathbf H} \ , \label{eq2} \\
\dot{\mathbf B}_{\rm rev} &=& -\nabla \times {\mathbf E} \ ,
\label{eq3}
\end{eqnarray}
\end{subequations}
\asja{in accord with \eqref{macro}.}

\asja{We note here that the lower-right $2\times 2$ block
matrix in $\mathbf L$, Eq.\eqref{L-matrix}, describing the
electromagnetic part of the system, has been used previously by
Marsden and Ratiu in \cite{Marsden}. Therein, it was given
in terms of a Poisson bracket at which we can arrive by using
$\left\{F,G \right\}= \int (\delta F/\delta {\mathbf x})\cdot
{\mathbf L} \cdot (\delta G/\delta {\mathbf x}) d^3r$.
In our work, their
Poisson bracket is embedded in a thermodynamic formulation with a
thermal variable $\varepsilon$.}

\asja{The degeneracy requirement and the antisymmetry of the
Poisson operator are satisfied by construction. For the Jacobi
identity, it is useful to observe that in our system the time
evolution of the entropy density $s\left( \varepsilon,{\mathbf
D},{\mathbf B} \right)$ has no reversible contributions. Since the
transformation of variables does not affect the Jacobi identity,
we choose to prove the Jacobi identity in the set of
variables ${\mathbf x'}=\left( s,{\mathbf D},{\mathbf B} \right)$,
in which the Poisson operator $\mathbf L'$ takes a much simpler
form. Namely, only two elements, ${L'}_{{\mathbf D}{\mathbf B}}$
and ${L'}_{{\mathbf B}{\mathbf D}}$, are non-zero, and $\mathbf
L'$ coincides with the Poisson bracket in \cite{Marsden}, for
which the Jacobi identity is fulfilled. With this, the Poisson
operator $\mathbf L$ is completely constructed.}

\subsection{Friction matrix obtained from microscopic fluctuations} \label{sec.3.d}

Coarse graining with respect to time from one level of description
(L1) to another one (L2) in principle leads to two kinds of
irreversible effects on the coarser level. First, dissipative
effects on level L1 reappear on level L2. Second, certain effects
that are slower than the time resolution on level L1 \asja{and
faster than the time resolution of L2} will be expressed as rapid
fluctuations on level L2, and hence emerge as irreversible effects
on that coarser level of description. \asja{It is this latter
kind of irreversible effects after coarse-graining which we
concentrate on. The coarse-graining procedure is captured in the
following five steps and illustrated for the case of dissipative
electromagnetism.}

\asja{First, we express the friction
matrix $\mathbf M$, which captures
the dissipative contributions to the GENERIC evolution equation
\eqref{generic}, in terms of a Green-Kubo-type expression. The
matrix $\mathbf M$ can be calculated from the microscopic
description by a generalization of the Green-Kubo relation
for the viscosity \eqref{viscosity}.}
If ${\dot {\mathbf x}}^{\rm f}$ denotes
the fluctuations on all components of ${\mathbf x}$, we write
\cite{hco},
\begin{eqnarray}\label{GreenKubo}
{\mathbf M}\left({\mathbf r},{\mathbf r'}\right) &=&
\frac{1}{k_{\rm B}}\int^\tau_0 dt \langle  \dot {\mathbf x}^{\rm
f}({\mathbf r}, t) \dot {\mathbf x}^{{\rm f}, T}({\mathbf r}',0)
\rangle \ ,
\end{eqnarray}
which can be expressed alternatively, in component notation, as
\begin{eqnarray}\label{epsilonGreenKubo}
\left[ 1+ \varepsilon (x_i)\varepsilon (x_k)
\right]M_{ik}\left({\mathbf r},{\mathbf r'}\right)=\frac{1}{k_{\rm
B} \tau} \langle \triangle x_i^{\rm f}({\mathbf r}) \triangle
x_k^{\rm f}({\mathbf r}') \rangle \ ,
\end{eqnarray}
with $\varepsilon (x_j)=\pm 1$ depending on the parity of $x_j$
under time-reversal. Here, $\tau$ is an intermediate time scale
separating the slow degrees of freedom \asja{(on the coarser level
L2)} from the fast ones \asja{(on the finer level L1)}. The
quantity $\triangle x_j^{\rm f}({\mathbf r})$ is given by
$\triangle x_j^{\rm f}({\mathbf r}) = \int^\tau_0 \dot x_j^{\rm
f}({\mathbf r}, t) \, dt$, and the brackets $\langle \ldots
\rangle$ indicate the average over an ensemble of microscopic
trajectories consistent with the slow macrostate $\mathbf x$ over
the whole time interval, see pages 357--358 in \cite{hco} for more
details. The superscript $T$ in \eqref{GreenKubo} denotes the
usual matrix transpose.

\asja{Second, for formulating the form of the fluctuations
${\dot {\mathbf x}}^{\rm f}$ the type of the variables ${\mathbf x}$ is
important. Specifically, choosing densities of (conserved) extensive
variables is particularly useful. Doing so is in close analogy to the
common procedure in the theory of fluctuations in classical
thermodynamics} \cite{Groot, Callen}. \asja{Furthermore, the Green-Kubo
relation \eqref{viscosity}, and similar relations for the thermal
conductivity and diffusion coefficient, indicate that current-current
correlations are quantities of fundamental importance. For example, the
stress tensor is the current of momentum. Current densities play a
major role in the formulation of conservation laws in the form
of field equations for density variables of extensive quantities, e.g.,
internal energy density, momentum density. Hence, from this
perspective it is tempting to use density variables in the set
${\mathbf x}$. How this can be exploited further is discussed in the
third step below. In our example, the energy density $\varepsilon$ is
hence a good variable, in contrast to the temperature $T$, or the
(not conserved) entropy density $s$. Further
dissipative processes we are considering here originate from the
fluctuations in the particle positions and spins. Thus,
fluctuations in $\mathbf{D}$ and $\mathbf{B}$ arise due to
fluctuations in densities of extensive variables, namely in the
polarization $\mathbf{P}$ and the magnetization $\mathbf{M}$.}

\asja{Third, we make an {\it ansatz} for the structural form of the
fluctuations. We assume} that the evolution equations for the
fluctuations are similar in structure to their macroscopic
counterparts, \asja{in the spirit of Onsager's regression
hypothesis. This hypothesis states that the fluctuations about the
equilibrium state decay, on the average, according to the same
laws that govern the decay of macroscopic deviations from
equilibrium} \cite{Kubo,Evans,hco}. In view of the first component
of $\dot{{\mathbf x}}^{\rm f}$ we note that
\eqref{energy_equation} represents the change in electromagnetic
energy only, while the corresponding equation for the thermal
energy, $\varepsilon_{\rm th}$, of the body will be of the form
$\dot \varepsilon_{\rm th} = {\mathbf J}_e\cdot {\mathbf E}
-\nabla \cdot \mathbf{J}_q$ with heat flux $\mathbf{J}_q$, such
that the total energy, i.e.\ the integral of
$\varepsilon=\varepsilon_{\rm em}+\varepsilon_{\rm th}$, is
conserved. Approximating the fluctuating contributions by white
noise leads to expressions for $\Delta {\mathbf x}^{\rm f}$ of the
form
\begin{eqnarray}\label{ansatz_Wiener}
\Delta {\mathbf x}^{\rm f}= \left(
\begin{array}{c}
- \nabla \cdot \left( {\mathbf W}^{({\rm q})}+ {\mathbf E}\times
{\mathbf W}^{({\rm H})}+{\mathbf W}^{({\rm E})}\times {\mathbf H}
\right)
\\
\nabla \times {\mathbf W}^{({\rm H})} - {\mathbf W}^{({\rm j})}
\\
- \nabla \times {\mathbf W}^{({\rm E})}
\end{array}
\right) \ ,
\end{eqnarray}
where all ${\mathbf W}$ are (small) increments of Wiener
processes, \asja{which are integrals of white noise} over a time
interval $\tau$. The superscripts to the Wiener processes indicate their
physical origin.

Fourth, by specifying the second moments of the Wiener
processes, and all correlations between them, the construction of
the $\mathbf M$ matrix is completed. We assume that all ${\mathbf
W}$ are decorrelated because of the difference in the underlying
microscopic mechanisms, but that ${\mathbf W}^{\rm q}$ has a
non-zero correlation only with the electric current fluctuations
${\mathbf W}^{\rm j}$, which will give rise to the thermoelectric
effects. Hence we write the following relations
%\begin{subequations} \label{correlations}
%\begin{eqnarray}
%\langle {\mathbf W}^{\rm q}({\mathbf r})  {\mathbf W}^{\rm
%q}({\mathbf r'}) \rangle &=& 2 k_{\rm B} \tau
%\delta({\mathbf r}-{\mathbf r}') {\mathbf C}^{\rm q}({\mathbf r}),\\
%\langle  {\mathbf W}^{\rm H}({\mathbf r})  {\mathbf W}^{\rm
%H}({\mathbf r'}) \rangle &=& 2 k_{\rm B} \tau
%\delta({\mathbf r}-{\mathbf r}') {\mathbf C}^{\rm H}({\mathbf r}),\\
%\langle  {\mathbf W}^{\rm j}({\mathbf r})  {\mathbf W}^{\rm
%j}({\mathbf r'}) \rangle &=& 2 k_{\rm B} \tau
%\delta({\mathbf r}-{\mathbf r}') {\mathbf C}^{\rm j}({\mathbf r}),\\
%\langle  {\mathbf W}^{\rm E}({\mathbf r}) {\mathbf W}^{\rm
%E}({\mathbf r'}) \rangle &=& 2 k_{\rm B} \tau
%\delta({\mathbf r}-{\mathbf r}') {\mathbf C}^{\rm E}({\mathbf r}),\\
%\langle  {\mathbf W}^{\rm q}({\mathbf r})  {\mathbf W}^{\rm
%j}({\mathbf r'}) \rangle &=& 2 k_{\rm B} \tau \delta({\mathbf
%r}-{\mathbf r}') {\mathbf C}^{\rm te}({\mathbf r}),
%\end{eqnarray}
%\end{subequations}
\begin{eqnarray}\label{correlations}
\langle  {\mathbf W}^{(\alpha)}({\mathbf r})  {\mathbf
W}^{(\beta)}({\mathbf r'}) \rangle &=& 2 k_{\rm B} \tau
\delta({\mathbf r}-{\mathbf r}') {\mathbf C}^{(\alpha
\beta)}({\mathbf r}),
\end{eqnarray}
\asja{with $\alpha, \beta \in \{{\rm q}, {\rm j}, {\rm E}, {\rm
  H}\}$, where the non-zero ${\mathbf C}^{(\alpha \beta)}$, i.e.,
  ${\mathbf C}^{({\rm q}{\rm q})}$, ${\mathbf C}^{({\rm j}{\rm
j})}$, ${\mathbf C}^{({\rm E}{\rm E})}$, ${\mathbf C}^{({\rm
H}{\rm H})}$ and ${\mathbf C}^{({\rm q}{\rm j})}\equiv{\mathbf
C}^{({\rm j}{\rm q}),T}$, represent the correlation functions,
with implicit $\mathbf r$-dependence, e.g., through temperature.
Note that we have assumed only local correlations between the Wiener
processes, i.e., between the fluctuations. Hence, we are dealing with a
  local field formulation. However, non-local effects can be included
if so desired.}

\asja{In the fifth and final step, the firction matrix ${\mathbf M}$ is
  calculated.
Using the fluctuations \eqref{ansatz_Wiener} in conjunction
with the correlations \eqref{correlations} we calculate the Green-Kubo
relations
for the elements of the matrix ${\mathbf M}({\mathbf r},{\mathbf
r}')$ by way of \eqref{epsilonGreenKubo}.} The final result for the
friction matrix is
\begin{subequations} \label{M-matrix}
\begin{eqnarray}
{\mathbf M}({\mathbf r},{\mathbf r}') = \left(
\begin{array}{ccc}
M_{\varepsilon \varepsilon} & M_{\varepsilon{\mathbf D}}&
M_{\varepsilon{\mathbf B}}
\\
M_{{\mathbf D}\varepsilon} & M_{{\mathbf D}{\mathbf D}}& 0
\\
M_{{\mathbf B}\varepsilon}& 0 & M_{{\mathbf B}{\mathbf B}}
\end{array}
\right),
\end{eqnarray}
with
\begin{eqnarray}
M_{\varepsilon \varepsilon}&=& {\mathbf \nabla}\cdot {\mathbf
C}^{({\rm q}{\rm q})}({\mathbf r}) \cdot \left({\mathbf \nabla
'}\delta({\mathbf r} -{\mathbf r'})\right)+ \nabla \cdot {\mathbf
E}({\mathbf r})\times {\mathbf C}^{({\rm H}{\rm H})}({\mathbf r})
\cdot \left({\mathbf \nabla '} \times {\mathbf E}({\mathbf
r'})\delta({\mathbf r} -{\mathbf r'})\right) \nonumber
\\
&&+ \nabla \cdot {\mathbf H}({\mathbf r})\times {\mathbf C}^{({\rm
E}{\rm E})}({\mathbf r}) \cdot \left({\mathbf \nabla '} \times
{\mathbf H}({\mathbf r'})\delta({\mathbf r} -{\mathbf r'})\right)
,
\\
M_{{\mathbf D}{\mathbf D}}&=&-{\mathbf \nabla}\times {\mathbf
C}^{({\rm H}{\rm H})}({\mathbf r})\cdot \left({\mathbf \nabla '}
\delta({\mathbf r} -{\mathbf r'})\right) \times \ + \
\delta({\mathbf r}-{\mathbf r'}){\mathbf C}^{({\rm j}{\rm
j})}({\mathbf r}) \ ,
\\
M_{{\mathbf B}{\mathbf B}}&=&-{\mathbf \nabla}\times {\mathbf
C}^{({\rm E}{\rm E})}({\mathbf r})\cdot \left({\mathbf \nabla '}
\delta({\mathbf r}-{\mathbf r'})\right) \times \ , \\
M_{\varepsilon{\mathbf D}}&=&{\mathbf \nabla} \cdot {\mathbf
C}^{({\rm q}{\rm j})}({\mathbf r}) \delta({\mathbf r}-{\mathbf
r}')+\nabla \cdot {\mathbf E}({\mathbf r})\times {\mathbf
C}^{({\rm H}{\rm H})}({\mathbf r}) \cdot \left({\mathbf \nabla
'}\delta({\mathbf r} -{\mathbf
r'})\right)\times  \ , \\
M_{{\mathbf D}\varepsilon}&=& {\mathbf C}^{({\rm q}{\rm
j}),T}({\mathbf r})\cdot \left( {\mathbf \nabla '} \delta({\mathbf
r}-{\mathbf r}')\right) - \nabla \times {\mathbf C}^{({\rm H}{\rm
H})}({\mathbf r}) \cdot \left({\mathbf \nabla '} \times {\mathbf
E}({\mathbf
r'})\delta({\mathbf r} -{\mathbf r'})\right) ,\\
M_{\varepsilon{\mathbf B}}&=& \nabla \cdot {\mathbf H}({\mathbf
r})\times {\mathbf C}^{({\rm E}{\rm E})}({\mathbf r}) \cdot
\left({\mathbf \nabla '}\delta({\mathbf r} -{\mathbf
r'})\right)\times  \ , \\
M_{{\mathbf B}\varepsilon}&=& - \nabla \times {\mathbf C}^{({\rm
E}{\rm E})}({\mathbf r}) \cdot \left({\mathbf \nabla '} \times
{\mathbf H}({\mathbf r'})\delta({\mathbf r} -{\mathbf r'})\right)
,
\end{eqnarray}
\end{subequations}
where it is understood that contractions $\cdot$ and cross
products $\times$ are executed in the order of occurrence from the
right to the left.

Given the expression \eqref{M-matrix} for
$\mathbf{M}$, it is straightforward to show that the degeneracy
condition ${\mathbf M} \cdot ({\delta E}/{\delta \mathbf x}) =
\mathbf{0}$ is indeed satisfied. \asja{The origin of this result goes
  back to the {\it ansatz} for the fluctuations
$\Delta {\mathbf x}^{\rm f}$ in \eqref{ansatz_Wiener}. Their specific
form ensures that the fluctuations occur on a submanifold of constant
energy, as can be shown readily by}
\begin{eqnarray}\label{energy_conservation}
\int \Delta {\mathbf x}^{{\rm f}, T} \cdot \frac{\delta E}{\delta
{\mathbf x}}\; d^3r = 0 \, ,
\end{eqnarray}
\asja{as required by the GENERIC. Equation \eqref{energy_conservation} is fulfilled already by writing the first component
of $\dot{{\mathbf x}}^{\rm f}$, namely the fluctuations in the
total energy density, as in \eqref{ansatz_Wiener}, which requires
it to be the divergence of a vector field. It is then
straightforward to show that \eqref{energy_conservation} is
satisfied if boundary terms can be neglected.}

If dissipative matrix $\mathbf M({\mathbf r},{\mathbf r}' )$ is
applied to the entropy gradient $\delta S/\delta {\mathbf
x}(\mathbf{r'})$, including an integration over ${\mathbf r}'$,
one gets the irreversible contributions to the time evolution equations.
Using the definitions
\begin{subequations}
\begin{eqnarray}
{\mathbf E}^*&=& \frac{1}{T}{\mathbf C}^{({\rm E}{\rm E})} \cdot
\left(
{\mathbf \nabla} \times {\mathbf H} \right) , \\
{\mathbf H}^*&=& - \frac{1}{T}{\mathbf C}^{({\rm H}{\rm H})} \cdot
\left(
{\mathbf \nabla} \times {\mathbf E} \right) , \\
{\mathbf J}^*_e&=& {\mathbf C}^{({\rm j}{\rm j})} \cdot
\frac{{\mathbf E}}{T}+ {\mathbf C}^{({\rm q}{\rm j}),T} \cdot
{\mathbf \nabla }\frac{1}{T} \ ,
\end{eqnarray}
\end{subequations}
one obtains for the irreversible part of the time evolutions of
$\mathbf{D}$ and $\mathbf{B}$
\begin{subequations}\label{irrev_evolution}
\begin{eqnarray}
\dot{\mathbf D}_{\rm irr}&=& {\mathbf \nabla} \times {\mathbf H}^*
- {\mathbf J}^*_e \ , \label{ieq1}\\
\dot{\mathbf B}_{\rm irr}&=& -\nabla \times {\mathbf E}^*
\label{ieq2},
\end{eqnarray}
\end{subequations}
where all functions are evaluated at ${\mathbf r}$. \asja{We
recognize in $\dot{\mathbf D}_{\rm irr}$ the Ohmic conduction
${\mathbf C}^{({\rm j}{\rm j})}$, while ${\mathbf{C}^{({\rm q}{\rm
j})}}$ represents the thermoelectric coupling \cite{Groot}.}

\subsection{Final set of evolution equations} \label{sec.3.e}

\asja{The final electromagnetic time evolution equations can be
conveniently written in the following way.} Combining the
reversible \eqref{rev_evolution} and irreversible
\eqref{irrev_evolution} contributions to the time evolution
equations of the fields $\mathbf D$ and $\mathbf B$, the resulting
temporally coarse-grained Maxwell equations differ from
\eqref{macro} in that the fields ${\mathbf E}$, ${\mathbf H}$ and
${\mathbf J}_e$ in \eqref{macro} have to be replaced by ${\mathbf
E}+{\mathbf E}^*$, ${\mathbf H}+{\mathbf H}^*$ and ${\mathbf
J}_e+{\mathbf J}^*_e$, respectively. The additional fields
${\mathbf E}^*$, ${\mathbf H}^*$ and ${\mathbf J}^*_e$ give rise
to the irreversible contributions to $\dot{\mathbf D}$ and
$\dot{\mathbf B}$. They are proportional to the quantities $\nabla
\times {\mathbf H}$, $\nabla \times {\mathbf E}$, ${\mathbf E}$
and $\nabla T$, as expected from conditions \eqref{thd_forces} and
the discussion in Sec.~\ref{sec2}, and represent the
nonequilibrium forces that tend to restore equilibrium. We mention
that the full macroscopic Maxwell equations obtained here can be
used to derive expressions for the complex permittivity
$\mbox{\boldmath $\epsilon $}(\omega)$ and permeability
$\mbox{\boldmath$ \mu$}(\omega)$ when examined in the linear
response regime \cite{Liu94,Liu98}.

The frequently considered balance equation for the energy density
\asja{can be obtained from \eqref{eq1} and the result of the
application of $\mathbf M({\mathbf r},{\mathbf r}' )$ to the
entropy gradient $\delta S/\delta {\mathbf x}(\mathbf{r'})$,
namely,}
\begin{eqnarray} \label{energyeq}
\dot{\varepsilon}= -{\mathbf \nabla} \cdot \left( {\mathbf
E}\times {\mathbf H} \right)-{\mathbf \nabla} \cdot \left({\mathbf
E}\times {\mathbf H}^* +{\mathbf E}^*\times {\mathbf H} + {\mathbf
C}^{({\rm q}{\rm q})} \cdot {\mathbf \nabla}\frac{1}{T}+{\mathbf
C}^{({\rm q}{\rm j})}\cdot \frac{{\mathbf E}}{T} \right).
\end{eqnarray}
The first term on the right side, which is the divergence of the
well known Poynting vector, represents the reversible part of the
time evolution equation, while the second term is the irreversible
part, describing the heat conduction \asja{(through ${\mathbf
C}^{({\rm q}{\rm q})}$)} and the Peltier effect \asja{(through
${\mathbf C}^{({\rm q}{\rm j})}$)}, as well as a modification of
the electromagnetic energy flux, i.e., of the Poynting vector, due
to the additional fields ${\mathbf E}^*$ and ${\mathbf H}^*$. This
modification of the energy flux was found also by Liu
\cite{Liu95,Liu99}.

For the time evolution of the entropy density $s$ one can write,
using the chain rule and rearranging the expression,
\begin{align}\label{entropy_evolution}
\dot{s}&= -\nabla \cdot \left( \frac{1}{T}{\mathbf C}^{({\rm
q}{\rm q})}\cdot {\mathbf \nabla}\frac{1}{T} + \frac{1}{T}{\mathbf
C}^{({\rm q}{\rm j})}\cdot \frac{{\mathbf E}}{T} \right)\nonumber
\\ &+ \left(
\begin{array}{c}
{\mathbf \nabla} \frac{1}{T}\\
\frac{{\mathbf E}}{T}\\
\frac{1}{T}\left({\mathbf \nabla} \times {\mathbf E}\right)\\
\frac{1}{T}\left({\mathbf \nabla} \times {\mathbf H}\right)
\end{array}
\right)
\cdot
\left(
\begin{array}{cccc}
{\mathbf C}^{({\rm q}{\rm q})}&{\mathbf C}^{({\rm q}{\rm j})}&{\mathbf 0}&{\mathbf 0}\\
{\mathbf C}^{({\rm q}{\rm j}),T}&{\mathbf C}^{({\rm j}{\rm j})}&{\mathbf 0}&{\mathbf 0}\\
{\mathbf 0}&{\mathbf 0}&{\mathbf C}^{({\rm H}{\rm H})}&{\mathbf 0}\\
{\mathbf 0}&{\mathbf 0}&{\mathbf 0}&{\mathbf C}^{({\rm E}{\rm E})}
\end{array}
\right)
\cdot
\left(
\begin{array}{c}
{\mathbf \nabla} \frac{1}{T}\\
\frac{{\mathbf E}}{T}\\
\frac{1}{T}\left({\mathbf \nabla} \times {\mathbf E}\right)\\
\frac{1}{T}\left({\mathbf \nabla} \times {\mathbf H}\right)
\end{array}
\right).
\end{align}
The central matrix of the last term is positive semidefinite as
can be seen after inserting the microscopic expressions for the
correlation functions \eqref{correlations}, and therefore the
entropy production rate is indeed non-negative. The obtained
entropy production rate corresponds to the one proposed by
Liu  in \cite{Liu93,Liu94}.

\section{Discussion and conclusions} \label{sec5}

Coarse-grained \asja{dissipative} Maxwell equations of
electromagnetism have been obtained and examined from the
perspective of nonequilibrium thermodynamics by using the GENERIC
formalism. \asja{We have illustrated a scheme to relate the
dissipative effects on the macroscopic scale to the fast
microscopic fluctuations in the polarization and magnetization in
the spirit of a generalized fluctuation-dissipation theorem.} In
particular, the study is applicable also to materials having
nonlinear constitutive relations between $({\mathbf E},{\mathbf
B})$ and $({\mathbf D},{\mathbf H})$. It emerged clearly that the
difference between spatial and temporal coarse-graining is of
fundamental importance. On the one hand, the usual procedure of
averaging the microscopic Maxwell equations only spatially (or by
single-time ensembles averages) leads to terms equivalent to the
reversible contributions in the GENERIC formulation, $\mathbf{L}
\cdot (\delta E/\delta \mathbf{x})$. On the other hand,
coarse-graining the spatially averaged Maxwell equations further
with respect to time leads to new irreversible effects captured in
terms of $\mathbf{M} \cdot (\delta S/\delta \mathbf{x})$, as shown
here. The friction matrix $\mathbf M$ has been related to two-time
correlations of fast processes on the finer level of description.
Using the Green-Kubo expression \eqref{GreenKubo}, the
correlations of microscopic fluctuations give rise to dissipative
processes such as Ohmic currents, the thermoelectric effect, and
to other irreversible contributions to the electric and magnetic
fields.

\asja{In the above investigations, the following points have to
be highlighted. First, we presented an illustration of the
coarse-graining procedure in order to obtain the structure
of fluctuations and, in turn, the structure of the dissipative
effects in electromagnetism.
%, rather than an explicit derivation of
%those dissipative effects in detailed microscopic terms.
Second, the identification of current densities occurring in
$\dot{\mathbf x}^{\rm f}$ was of fundamental importance, since
correlations between them lead to the new dissipative processes on
the macroscopic scale. To aid the relation between the
Green-Kubo-type expression for ${\mathbf M}$ \eqref{GreenKubo} and
the correlations between current densities, it has proven useful
to choose density variables of extensive quantities in the set
${\mathbf x}$. Using Onsager's regression hypothesis, the
evolution equations of the fluctuations takes the form of
conservation laws, including fluctuating current densities. We
point out that the correlations \eqref{correlations} are similar
in spirit to the Green-Kubo relation \eqref{viscosity}, in the
sense that they define transport coefficients. In contrast, the
expression for ${\mathbf M}$ \eqref{GreenKubo} in conjunction with
the entropy gradient leads to the full form of the dissipative
processes. The third point concerns the details behind the
correlations \eqref{correlations}. In order to specify the
dissipative processes beyond their structure, one needs to discuss
in more detail the physics behind the correlation matrices
${\mathbf C}^{\alpha \beta}$, with $\alpha, \beta \in \{{\rm q},
{\rm j}, {\rm E}, {\rm H}\}$. That is, the transport coefficients
appearing in the final expressions through ${\mathbf C}^{\alpha
\beta}$ can be obtained, e.g., via molecular dynamics simulations,
or be derived based on explicit expressions for the microscopic
fluxes and the correlations between them \cite{Pep03}. }

\asja{The scheme of temporal coarse-graining for obtaining the
  macroscopic Maxwell equations differs from Liu's} \cite{Liu93,Liu94},
\asja{who started by making an {\it ansatz} for the entropy
production rate. However, both procedures lead to identical
  modifications of the Maxwell equations, and also the energy equation
 \eqref{energyeq} is in agreement with the
corresponding relations in} \cite{Liu93,Liu94}. \asja{The obtained
Maxwell relations and extensions of them are successfully applied
to the problems of nematic liquid crystals} \cite{Liu93,Liu94},
\asja{colloidal magnetic and electric fluid, and ferrofluids}
\cite{Liu95,Liu98,Liu99}. \asja{In nematic liquid crystals, the
dissipative couplings between the new thermodynamic forces $\nabla
\times {\mathbf E}$ and $\nabla \times {\mathbf H}$ with the
additional thermodynamic forces specific for this system are
permitted and are equipped with transport coefficients, thus being
valid also in case of the oscillatory instabilities, when usually
employed static Maxwell equations no longer hold}
\cite{Liu93,Liu94}. \asja{Liu's equations for ferrofluids also
give rise to the dissipative forces that account for the special
spin-up behavior of ferrofluids under a rotating external field}
\cite{Liu95,Liu98,Liu99}.

A cross relationship, similar to the one suggested by Liu
\cite{Liu93}, occurred in the dissipative terms
\eqref{irrev_evolution} obtained by us: ${\mathbf C}^{({\rm H}{\rm
H})}$ in $\dot{\mathbf D}_{\rm irr}$ accounts for the correlations
of the fluctuations in magnetization, i.e.\ the magnetization
relaxation time, and ${\mathbf C}^{({\rm E}{\rm E})}$ in
$\dot{\mathbf B}_{\rm irr}$ accounts for the correlations of the
fluctuations in polarization, i.e.\ the polarization relaxation
time. This is the consequence of the nature of Maxwell equations
and of the mutual dependence of the electric and magnetic fields.
It is naturally obtained by the form \eqref{ansatz_Wiener} for the
increments in fluxes $\Delta {\mathbf x}^{\rm f}$.

Aspects, different from the
ones discussed here, of dissipative effects in electromagnetism and
of linking electromagnetism with hydrodynamics can be found in
\cite{Feld99, Groot, Feld04}.
Decomposition of the electromagnetic fields, in particular polarization
and magnetization, into the slow and fast
parts has been considered by Felderhof and Kroh
\cite{Feld99}. In this way they extended the irreversible
thermodynamics approach used by de Groot and Mazur \cite{Groot}
for the hydrodynamics of magnetic and dielectric fluids in
interaction with the electromagnetic field.
In a subsequent paper \cite{Feld04}, Felderhof studied
semirelativistic hydrodynamic evolution equations for a medium with
polarization and magnetization.

%%%%%%%%%%%%%%%%%%%%%%%%%%%%%%%%%%%%%%%%%%%%%%%%%%%%%%%%%%%%%%%%%%

\end{document}